\documentclass[a4paper,11pt]{article}
\pdfoutput=1
\usepackage{jcappub}
\usepackage{fontenc}

\usepackage{units}
\usepackage{xspace}
\usepackage{subfig}
\usepackage{epstopdf}
\usepackage{units}
\usepackage{graphicx}
\usepackage{dcolumn}
\usepackage{bm}
\usepackage{epsfig}
\usepackage{url}
\usepackage{aas_macros}
\usepackage{afterpage}

\allowdisplaybreaks

\newcommand*{\los}{line-of-sight\xspace}
\newcommand*{\SONG}{\textsf{SONG}\xspace}

\newcommand*{\lmax} {\ensuremath{\ell_\text{max}}\xspace}

\newcommand*{\fnl}{\ensuremath{f_{\text{NL}}}\xspace}
\newcommand*{\sci} [2] {\ensuremath{{#1} \times 10^{#2}}} 
\newcommand*{\ie} {i.\,e.\xspace}
\newcommand*{\eg} {e.\,g.\xspace}

\newcommand*{\avg} [1] {\ensuremath{\left\langle\,{#1}\,\right\rangle}\xspace}

\newcommand{\dd}{\textrm{d}}
\renewcommand{\L}{\ensuremath{\ell}\xspace}
\newcommand{\lm}{\ensuremath{\ell m}\xspace}
\newcommand{\lmp}{\ensuremath{\ell' m'}\xspace}
\newcommand*{\msk}{\\[0.25cm]} 
\newcommand*{\nmsk}{\notag\msk} 
\newcommand{\kone}{\ensuremath{\vec{k_1}}\xspace}
\newcommand{\ktwo}{\ensuremath{\vec{k_2}}\xspace}

\newcommand{\pert}[2]{\ensuremath{#1^{(#2)}}\xspace}

\defcitealias{mollerach:2004a}{MHM2004}
\newcommand*{\MHM}{\citetalias{mollerach:2004a}\xspace}

\defcitealias{beneke:2011a}{BFK2011}
\newcommand*{\BF}{\citetalias{beneke:2011a}\xspace}

\renewcommand{\vec}[1]{{\mathbf {#1}}} 

\newcommand*{\sref} [1] {Section~\ref{#1}\xspace}

\newcommand*{\eref} [1] {Eq.~\eqref{#1}\xspace}

\newcommand*{\fref} [1] {Figure~\ref{#1}\xspace}

\title{The intrinsic B-mode polarisation of the Cosmic Microwave Background}
\author[1]{Christian Fidler,}
\author[1,2]{Guido W.\ Pettinari,}
\author[3]{Martin Beneke,}
\author[1]{Robert Crittenden,}
\author[1]{Kazuya Koyama}
\author{and}
\author[1]{David Wands}
\affiliation[1]{Institute of Cosmology and Gravitation, University of
Portsmouth, Portsmouth PO1 3FX, UK}
\affiliation[2]{Department of Physics \& Astronomy, University of Sussex,
Brighton BN1 9QH, UK}
\affiliation[3]{Physik Department T31, Technische 
  Universit\"at M\"unchen, D -- 85748 Garching, Germany}
\emailAdd{Christian.Fidler@port.ac.uk}
\emailAdd{Guido.Pettinari@sussex.ac.uk}
\emailAdd{Robert.Crittenden@port.ac.uk}
\emailAdd{Kazuya.Koyama@port.ac.uk}
\emailAdd{David.Wands@port.ac.uk}

\abstract{
We estimate the B-polarisation induced in the Cosmic Microwave Background by the non-linear
evolution of density perturbations. Using the second-order Boltzmann code SONG, our analysis incorporates, for the first time, all physical effects at recombination.
We also include novel contributions from the redshift part of the Boltzmann equation and from the bolometric definition of the temperature in the presence of polarisation. The remaining line-of-sight terms (lensing and time-delay) have previously been studied and must be calculated non-perturbatively.
The intrinsic B-mode polarisation is present independent of the initial conditions
and might contaminate the signal from primordial gravitational
waves.
We find this contamination to be comparable to a primordial tensor-to-scalar
ratio of $r\simeq10^{-7}$ at the angular scale $\ell\simeq100\,$, where the
primordial signal peaks, and $r\simeq\sci{5}{-5}$ at $\ell\simeq700\,$, where the
intrinsic signal peaks.
Therefore, we conclude that the intrinsic B-polarisation from second-order
effects is not likely to contaminate future searches of primordial gravitational
waves.
}

\begin{document}
{\raggedright\mbox{TUM-HEP-924/14}}
\maketitle
\flushbottom


\section{Introduction}
\label{sec:introduction}

The polarisation of the cosmic microwave background (CMB) encodes important
information about the origin of the primordial perturbations,
as well as their subsequent evolution.
The linear polarisation of the CMB is usually described in terms of its
curl-free and gradient-free components, the E and B polarisation modes.
The E-polarisation has been detected \cite{kovac:2002a, bennett:2003a} and is
routinely used jointly with the CMB temperature fluctuations to constrain the
cosmological parameters and to break their degeneracies \cite{hinshaw:2012a,
bennett:2012a, planck-collaboration:2013a}.
At leading order, the B-polarisation of the CMB is only induced by primordial
gravitational waves, \ie~the tensor modes of the metric.
The mechanism of cosmic inflation naturally generates a stochastic background of
primordial gravitational waves; its amplitude is proportional to the
energy scale of inflation. Therefore, measuring the primordial B-polarisation
would directly constrain models of inflation and provide an indirect
detection of the gravitational waves \cite{kamionkowski:1997a, seljak:1997a,
hu:1997b}.

The amplitude of the gravitational waves background produced in the early Universe is
usually parametrised via the tensor-to-scalar ratio $r\,$, the ratio in
power between the primordial tensor and scalar perturbations.
The current CMB observations are consistent with the absence of primordial
B-polarisation and translate to an upper limit on the tensor-to-scalar ratio of
order $r<\mathcal{O}(10^{-1})\,$, as found by both space-based (WMAP
\cite{bennett:2012a}) and ground-based experiments (QuAD \cite{brown:2009a},
QUIET \cite{quiet-collaboration:2012a}, BICEP1 \cite{bicep1-collaboration:2013a}).
A huge effort is currently being made to design CMB polarisation experiments
that significantly improve these constraints \cite{crill:2008a, sheehy:2010a, kogut:2011a,
austermann:2012a, staniszewski:2012a, prism-collaboration:2013b}.
In particular, the LiteBIRD \cite{hazumi:2012a,matsumura:2013a}, PIXIE
\cite{kogut:2011a} and PRISM \cite{prism-collaboration:2013b} space-borne experiments promise to determine $r$ with a
precision of $\Delta\,r = \mathcal{O}(10^{-3})\,$, $\mathcal{O}(10^{-3})\,$  and $\mathcal{O}(10^{-4})$, respectively.

There are, however, a number of non-linear effects that induce some degree of
B-polarisation even in absence of primordial gravitational waves. This happens
because, at the non-linear level, scalar, vector and tensor perturbations naturally
couple and source each other; thus, tensor perturbations always arise due to the
non-linear evolution of the primordial scalar fluctuations.
This mechanism could bias
the measurement of the B-polarisation of primordial origin, especially if $r$ is
as small as the detection threshold of PRISM.
B-polarisation from non-linear effects cannot be treated with linear perturbation theory
\cite{kamionkowski:1997a, seljak:1997a, hu:1997b}: either second-order perturbation theory or a non-perturbative
approach is needed to correctly account for them.

The most important contamination is weak lensing, which converts
E-polarisation to B-polarisation as the CMB photons travel through the inhomogeneous
universe \cite{zaldarriaga:1998c}.
Weak lensing becomes large at small scales and late times, when perturbation
theory breaks down; the usual treatment of weak lensing therefore avoids
perturbation theory by considering the small deflection angles of the photon
trajectories \cite{lewis:2006a}. 
The B-polarisation from lensing limits a detection of tensor modes to
$r>\mathcal{O}(10^{-3})$ \cite{lewis:2002a, knox:2002a, hu:2002b}; however
delensing techniques can be used to bring this limit to $r>\mathcal{O}(10^{-6})$
in absence of a sky cut \cite{seljak:2004a, lewis:2006a, smith:2009a}.
These B-modes generated by lensing have been detected in
the CMB by the SPTpol experiment at $\sim8\,\sigma$ \cite{hanson:2013a}.
Another non-perturbative effect that accumulates along the photon paths is the
gravitational time-delay. Similarly to lensing, time-delay converts E-modes to
B-modes, but its amplitude is strongly suppressed with respect to lensing for
geometrical reasons \cite{hu:2001a}. As a result, the time-delay bias on a measurement of
the primordial tensor modes is smaller than the detection threshold of a PRISM-like
experiment.

The remaining sources of B-polarisation from non-linear evolution can be
treated using second-order perturbation theory.
The Einstein and Boltzmann equations at second order have been studied in great
detail \cite{bartolo:2006a, bartolo:2007a, pitrou:2007a, pitrou:2009b,
beneke:2010a, naruko:2013a}. They are significantly more complicated than at
first order and solving them numerically is a daunting task.
Mollerach, Harari and Matarrese (2004) \cite{mollerach:2004a} (hereafter \MHM)
simplified the problem by studying only the linear response of the CMB photons to the
second-order vector and tensor perturbations, which are known analytically \cite{lu:2009a,
christopherson:2009a, christopherson:2011a, christopherson:2011b, ananda:2007a,
osano:2007a, baumann:2007a, arroja:2009a}; by doing so, they found that the
B-modes from this non-linear effect dominate over the primordial ones unless the
tensor-to-scalar ratio is of order $r>\mathcal{O}(10^{-6})\,$.
The complementary effect of the second-order collisions between photons and
electrons during recombination has been computed by Beneke, Fidler and Klingm\"uller (2011)
\cite{beneke:2011a} (hereafter \BF) using a full second-order approach. They
found the effect to be small compared to the lensing contribution, but of
comparable order to that of the metric sources considered by \MHM.

In this paper we use the second-order Boltzmann code \SONG
\cite{pettinari:2013a} to compute the power spectrum of the B-polarisation from
non-linear evolution.
With respect to the perturbative studies described above, we adopt a more general treatment whereby:
\begin{itemize}
  \item We consider all the second-order sources generating B-polarisation at the time of recombination and take into account their correlation. These include the Liouville sources, which we compute for the first time, and the collision and metric sources.
  \item We compute for the first time the B-mode polarisation generated by the redshift term in the Boltzmann equation along the line-of-sight.
  \item We compute the bolometric temperature and account for its implications on polarisation.
\end{itemize}
Our approach is the most complete to date and only lacks the late-time lensing and time delay effects. As we have discussed above, these effects have already been computed in the literature non-perturbatively.

\paragraph{Outline of the paper}
In section \ref{sec:boltzmann} we present the second-order polarised Boltzmann equations, establish our notation and discuss how B-modes are generated at second order. In section \ref{sec:line_of_sight} we introduce the line-of-sight formalism in order to compute the polarised transfer functions today. We also show that the polarised redshift terms in the Boltzmann equation can be absorbed by a transformation of variables, and provide a consistent definition of temperature including the implications for polarisation. Section \ref{sec:spectrum} summarises the computation of the B-mode power spectrum. Our results are presented in section \ref{sec:results}, where we also discuss how they relate to the previous literature and investigate the numerical stability of \SONG. Finally we report our conclusions is section \ref{sec:conclusions}.

\paragraph{Metric}
We employ the Poisson or conformal Newtonian gauge \cite{bertschinger:1995a, bruni:1997a},
where the metric is given by:
\begin{equation}
  \dd s^2 = a^2(\eta) \, \left\{ - (1+2\Psi) \dd \eta^2
    + 2 \omega_i \dd x^i \dd \eta
  + \,\left[\,(1-2\Phi) \delta_{ij} + 2\,\gamma_{ij} \,\right]\, \dd x^i \dd x^j
\right\} \;.
\label{eq:metric}
\end{equation}
The vector potential is transverse $\,\partial_i\,\omega_i=0\,$, while the
spatial metric is both transverse $\,\partial^i\,\gamma_{ij}=0\,$ and traceless
$\,{\gamma^i}_i=0\,$.

\paragraph{Cosmological parameters}
Throughout this paper we assume a flat $\Lambda$CDM model with Planck
parameters \cite{planck-collaboration:2013a}: $h = 0.678$, $\Omega_b =
0.0483$, $\Omega_{\text{cdm}} = 0.259$, $\Omega_\Lambda = 0.693$, $A_s =
\sci{2.214}{-9}$, $n_s = 0.961$, $\kappa_\text{reio} = 0.095$, $N_\text{eff} =
3.04$.
conformal age $\eta_0 = \unit[14210]{Mpc}$ and recombination happens at
$z=1089$, corresponding to a conformal time of $\eta_\text{rec} =
\unit[281]{Mpc}$. (The above parameters are given in natural units where $c=1$.) We assume adiabatic initial conditions with a vanishing primordial tensor-to-scalar ratio~($r=0$), and therefore focus only on the B-polarisation that is generated in the absence of primordial tensor perturbations.
We consider Gaussian initial conditions~($\fnl=0$) without loss of generality, because, for purely-scalar initial conditions, the contribution to the power spectrum from the three-point function is at least fifth-order in the primordial perturbations and thus negligible (\sref{sec:spectrum}).



\section{Boltzmann equation}
\label{sec:boltzmann}

The Boltzmann equation describes how the distribution function of a particle
species evolves in perturbed space-time, given its interactions with the other
species.
To describe polarised radiation in the Boltzmann formalism, one has to introduce
a Hermitian tensor-valued distribution function, $\,f_{\mu\nu}(\eta,\vec x, \vec
q)\,$, with $\vec{q}=q\vec{n}$ being the comoving momentum in a locally inertial frame, such that
\begin{align}
  \hat\epsilon^\mu\;\hat\epsilon^{*\nu}\;f_{\mu\nu}\,(\eta,\vec x, \vec q)
\end{align}
is the number density of photons at ($\vec x$,\,$\vec q$) in phase space and at time
$\eta$ in the polarisation state $\hat\epsilon$ (see
Refs.~\cite{pitrou:2009b,beneke:2010a,naruko:2013a} and references therein). 
The polarised distribution function can be decomposed on the so-called
\emph{helicity basis} of the spherical coordinate system,
\begin{align}
  f^{\mu\nu} \;=\; \sum\limits_{ab}\;f_{ab}\;
  {\vec\epsilon}^{\nu}_b\;{\vec\epsilon}^{*\mu}_a \;,
\end{align}
given by the two vectors
\begin{align}
  {\vec\epsilon}_+ \;=\; -\frac{1}{\sqrt{2}}\,(\,\vec e_\theta\,+\,i\,\vec
e_\phi\,)
  \qquad\text{and}\qquad
  {\vec\epsilon}_- \;=\; -\frac{1}{\sqrt{2}}\,(\,\vec e_\theta\,-\,i\,\vec
e_\phi\,) \;,
\end{align}
where $\theta$ and $\phi$ are the polar and azimuthal angles of the direction of propagation of the photon, $\,\vec
n$, while the vectors
$\,\vec e_\theta=\partial_\theta\vec n\,$ and $\,\vec
e_\phi=\partial_\phi\vec n/\sin\theta\,$
span the plane orthogonal to $\vec{n}$. The $a$ and $b$ indices are called helicity indices and can assume the values
$\,ab=++,\,--,\,-+,\allowbreak\,+-\,$. We follow the conventions used in Ref.~\cite{beneke:2010a} for the choice of coordinate system and the polarisation basis. 

The four physical degrees of freedom of $\,f_{ab}\,$ can also be expressed in
terms
of the Stokes parameters,
\begin{align}
  f_{ab} \;=\; 
  \left(\begin{array}{cc} f_{++} & f_{+-} \\ 
  f_{-+} & f_{--}\end{array}\right) \;=\; 
  \left(\begin{array}{cc} f_I-f_V & f_Q-if_U \\ 
  f_Q+if_U & f_I+f_V\end{array}\right) \;,
  \label{eq:Stokes}
\end{align}
where $f_I$ is the intensity, $f_V$ the circular polarisation, $f_Q$ and $f_U$
the two components of linear polarisation.
The intensity is related to the photon temperature; what we have been referring
to as $f$ in Ref.~\cite{pettinari:2013a} is, in the formalism of polarised
radiation, $f_I\,$.
We shall see below that $f_Q$ and $f_U$ are related to the E and B-polarisation
by multipole decomposition.
The circular polarisation, $f_V$, is not sourced by any mechanism in the
standard cosmological paradigm, and we shall therefore ignore it.

The time evolution of the photon distribution function $\,f_{ab}(\eta, \vec{x},
\vec{q})\,$ at second order is determined by the Boltzmann equation. Before
multipole decomposition, this simply reads
\begin{equation}
	\label{eq:boltz}
	\frac{\partial f_{ab}}{\partial \eta} + \frac{dx^i}{d\eta}\frac{\partial
f_{ab}}{\partial x_i} +
  \frac{dq}{d\eta}\frac{\partial f_{ab}}{\partial q}  +
  \frac{dn^i}{d\eta}\,
  \left(\frac{\partial f_{ab}}{\partial n_i}+\epsilon_{a\,j}\frac{\partial
\epsilon_c^{j*}}{\partial {n}^i} f_{cb}
    +\epsilon_{b\,j}^* \frac{\partial \epsilon_c^j }{\partial {n}^i}
f_{ac}\right)
  \,= \,\mathfrak{C}[f_{ab}] \;.
\end{equation}
The Liouville term on the left-hand side describes particle propagation in an
inhomogeneous space-time and the collision term on the right-hand side 
describes the interactions between photons and electrons through Compton
scattering. (Note that the $a,b,c$ indices refer to helicity, $a,b,c=+,-$, while
the other latin indices are spatial, $i,j,k=1,2,3$.)
The gradient term ($\frac{dx^i}{d\eta}\frac{\partial
f_{ab}}{\partial x_i}$) encodes free streaming; at higher order this term
also includes time-delay effects.
The momentum-derivative term ($\frac{dq}{d\eta}\frac{\partial f_{ab}}{\partial
q}$) causes the redshifting of photons at background level and at
higher-order includes the well-known Sachs-Wolfe (SW), integrated Sachs-Wolfe
(ISW) and Rees-Sciama (RS) effects.
The terms in $\frac{dn^i}{d\eta}$ vanish to first order and describe the
small-scale effect of gravitational lensing on the CMB.
The polarisation vectors appearing in these lensing terms encode the physical effect
whereby a change in the direction of propagation of the photon results into a rotation
of the polarisation basis. This is the second-order equivalent of the well known
non-perturbative effect \cite{zaldarriaga:1998c} that mixes different polarisation types.


It is convenient to express the Boltzmann equation using the brightness
$\Delta_{ab}$, defined as the momentum-integrated distribution function
normalised to the background energy density,
\begin{equation}
	\label{eq:Deltadef}
	\delta_{ab}+\Delta_{ab}(\eta,\vec{x},\vec{n}) \;=\; 
	\frac{\int dq \, q^3 \, f_{ab}(\eta,\vec{x},q\vec{n})}
	{\int dq \, q^3 \, f^{(0)}(q)} \;,
\end{equation} 
where $f^{(0)}$ is the background intensity distribution function, which takes the form of
a black-body spectrum.
The total time derivatives in the Liouville term can be evaluated using the
geodesic equation \cite{bartolo:2006a, pitrou:2009a,senatore:2009a,beneke:2010a}
in Newtonian gauge, yielding\footnote{Note that we have corrected a typographical error in the sign
of $n^i \dot{\omega}_i$ in Eq.~(2.3) of
Ref.~\cite{pettinari:2013a}.}
\begin{align}\nonumber
	&\frac{dx^i}{d\eta}\frac{\partial f_{ab}}{\partial x_i} &\rightarrow&
\quad
  (1 + \Psi + \Phi)\,n^i\partial_i\Delta_{ab} \;, \\ \nonumber
	&\frac{dq}{d\eta}\frac{\partial f_{ab}}{\partial q} &\rightarrow& \quad
    4\,(n^i\partial_i \Psi -\dot{\Phi} + n^i \dot{\omega}_i + n^i n^j
\dot{\gamma}_{ij})\,(\delta_{ab}
    + \Delta_{ab}) - 4 (\Psi -\Phi)\,n^i\partial_i \Psi\,\delta_{ab} - 8
\Phi\dot{\Phi}\,\delta_{ab} \;, \\
	&\frac{dn^i}{d\eta}\frac{\partial f_{ab}}{\partial n_i} &\rightarrow&
\quad 
    -(\delta^{ij} - n^i n^j)\,\partial_j(\Psi+\Phi)\, \left(\,\frac{\partial
\Delta_{ab}}{\partial n^i}+\,
    \epsilon_{a\,k}\frac{\partial \epsilon_c^{k*}}{\partial {n}^i} \Delta_{cb}
+\epsilon_{b\,k}^*
    \frac{\partial \epsilon_c^k}{\partial {n}^i} \Delta_{ac}\right) \,.
	\label{eq:timedelay_lensing_redshift}
\end{align}
We represent the terms containing only metric perturbations by 
\begin{equation}
\mathcal{M}_{ab} \,\,=\,\,  \delta_{ab} \;\left[\;4 \, (n^i\partial_i \Psi
-\dot{\Phi}
  + n^i \dot{\omega}_i + n^i n^j \dot{\gamma}_{ij})
  \,-\, 4 (\Psi -\Phi)\,n^i\partial_i \Psi \,-\, 8 \Phi\dot{\Phi} \;\right] \,,
  \label{eq:metric_terms}
\end{equation}
and the remaining quadratic terms, including the photon distribution
$\Delta$, by
\begin{multline}
  \label{eq:QL}
  \mathcal{Q}^L_{ab} \;=\; (\Psi + \Phi)\,n^i\partial_i\Delta_{ab} \;+\; 4 \,
    (n^i\partial_i \Psi -\dot{\Phi})\,\Delta_{ab} \\
    \;-\; (\delta^{ij} - n^i   n^j)\,
    \partial_j(\Psi+\Phi)\,\left(\,\frac{\partial \Delta_{ab}}{\partial n^i}+\,
    \epsilon_{a\,k}\frac{\partial \epsilon_c^{k*}}{\partial {n}^i} \Delta_{cb}
+\epsilon_{b\,k}^*
    \frac{\partial \epsilon_c^k}{\partial {n}^i} \Delta_{ac}\right) \;.
\end{multline}
The first term of $\mathcal{Q}^L$ contributes to the time-delay effect, while we
shall refer to the second and third ones as \emph{redshift} and \emph{lensing} terms, respectively.
The above expressions are valid only when $\Delta$ and the metric
variables are expanded up to second order $\Delta = \Delta^{(1)}+ \Delta^{(2)}$ and when $\omega_i^{(1)}=0$ and
$\gamma_{ij}^{(1)}=0\,$, that is, when the first-order initial perturbations are purely scalar.


We perform a Fourier transform from real space, ${\vec x}$, into comoving
wavevector, $\vec{k}$, and a decomposition from photon
direction, $\vec{n}$, into spherical harmonics, $(\ell,m)$. 
This transforms the Boltzmann and Einstein equations from partial differential
equations into hierarchies of ordinary differential equations.
The off-diagonal helicity components of the distribution function,
$f_{+-}=f_Q-if_U$ and $f_{-+}=f_Q+if_U$, transform under rotations around the
direction of photon propagation with spin $s=2$ and $-2$, respectively
\cite{lewis:2006a}.
Hence, they are decomposed into multipole space using spin-weighted spherical
harmonics \cite{kamionkowski:1997a, seljak:1997a, hu:1997b}, thus defining the
polarisation modes E and B:
\begin{eqnarray}
\label{eq:multipole_decomposition}
f_{E,\lm} \,\pm\, if_{B,\lm} &\,=\,& i^\L \sqrt{\frac{2\L+1}{4\pi}}\int d\Omega
\;
  Y_{\lm}^{ \mp 2 *}(\bm{n}) \; [\,f_Q(\bm{n}) \,\pm\, i f_U(\bm{n})\,] \,,\\
\nonumber
f_{I,\lm} &\,=\,& i^\L \sqrt{\frac{2\L+1}{4\pi}}\int d\Omega \;
  Y_{\lm}^*(\bm{n}) \, f_I(\bm{n}) \,,
\end{eqnarray}
where we have used the fact that $f_I=(f_{++}+f_{--})/2$
transforms with spin zero.
Note that while Q and U transform under rotations around the direction of
photon propagation, E and B are invariant by construction.
The E-polarisation has the same parity as the intensity field, $f_{E,\lm} \rightarrow (-1)^\ell\,f_{E,\lm}$, while the B-polarisation
has the opposite parity, $f_{B,\lm}\rightarrow(-1)^{\ell+1}\,f_{B,\lm}$.

We shall use a single composite index, $n$, to express the harmonic and
polarisation dependence of the distribution function, so that $\Delta_n$ will
denote $\Delta_{X,\lm}$ with $X=I,E,B$.
Then, the Boltzmann equation (\ref{eq:boltz}) and
(\ref{eq:timedelay_lensing_redshift}) simply reads
\begin{equation}
	\label{eq:boltzmann_compact}
	\dot{\Delta}_{n} \;+\; k\,\Sigma_{nn'}\Delta_{n'} \;+\; \mathcal{M}_{n} 
  \;+\; \mathcal{Q}^L_{n} \;=\; \mathfrak{C}_n \;,
\end{equation}
where $\Sigma_{nn'}$ is the free-streaming matrix that arises from the
decomposition of $n^i\partial_i\Delta$ into spherical harmonics. This term
couples neighbouring multipoles leading to the excitation of high-$\ell$ moments
over time and, as we shall see in \sref{sec:generation_mechanisms}, it also
couples E and B polarisation. 
The collision term
\begin{equation}
	\label{eq:collision_term}
	\mathfrak{C}_n \;=\; -|\dot{\kappa}|\,\left(\;\Delta_{n} \;-\;
    \Gamma_{nn'}\Delta_{n'} \;-\; \mathcal{Q}^C_{n}\,\,\right) \;,
\end{equation}
is proportional to the Compton scattering rate $|\dot{\kappa}|$ and consists of
three distinct contributions: the purely second-order loss term
$-|\dot{\kappa}|\,\Delta_{n}$, describing scatterings out of a given mode
$\Delta_n$, the purely second-order gain term
$|\dot{\kappa}|\,\Gamma_{nn'}\Delta_{n'}$, and quadratic contributions in $\Delta$ and in the electron velocity,
$|\dot{\kappa}|\,\mathcal{Q}^C_n$.
When the scattering rate is large, photons and baryons behave as a tightly
coupled fluid where the polarisation and all temperature moments larger than the quadrupole are suppressed. For the explicit expressions of the terms in Eq.~(\ref{eq:boltzmann_compact}) and (\ref{eq:collision_term}), refer to Eq.~(143) to (146) of Ref.~\cite{beneke:2010a}.
In order to connect this abstract notation to the well-known linear equations (see, for example, Eqs.~(61), (63) and (64) of Ref.~\cite{hu:1997b}), we report the purely second-order collision gain term:
\begin{align}
  &\Gamma_{(I,\lm)n'}\,\Delta_{n'} \;=\; \delta_{\ell0}\,\Delta_{I,00} \;
   +\;\delta_{\ell1}\,4\,v_e^{[m]} \;
   +\;\delta_{\ell2}\,(\,\Delta_{I,2m}\,-\,\sqrt{6}\,\Delta_{E,2m}\,)/10 \nmsk
  &\Gamma_{(E,\lm)n'}\,\Delta_{n'} \;=\; -\delta_{\ell2}\;\sqrt{6}\;
  (\,\Delta_{I,2m}\,-\,\sqrt{6}\,\Delta_{E,2m}\,)/10 \;, \nmsk
  &\Gamma_{(B,\lm)n'}\,\Delta_{n'} \;=\; 0 \;,
  \label{eq:compact_collision_term_gammamatrix}
\end{align}
where $v_e^{[m]}$ is the electron velocity and $m$ is the considered azimuthal mode.

\subsection{Generation mechanisms}
\label{sec:generation_mechanisms}


Only non-scalar sources can generate B-polarisation \cite{hu:1997b}.
In general, a perturbation $X(\vec{k})$ in Fourier space is a scalar if it is invariant under rotations around $\vec{k}$.
A second-order perturbation $\Delta^{(2)}(\vec{k})$ is sourced by the convolution of two first-order ones, e.g. $\Delta^{(1)}(\vec{k}_1)$ and $\Delta^{(1)}(\vec{k}_2)$.
Even if these are scalars with respect to $\vec{k_1}$ and $\vec{k_2}$, in general they are non-scalar with respect to $\vec{k}$, unless the three wave-vectors are aligned.
This means that the non-linear structure of the second-order equations naturally induces a variety of non-scalar sources even when the linear perturbations are scalar; these can in turn generate B-mode polarisation.
This is in sharp contrast to the first-order case, where non-scalar modes cannot be induced unless they are present in the initial conditions, as is the case for primordial gravitational waves.


We present the second-order Boltzmann equation for B-mode polarisation explicitly, and discuss its different sources. In this context, we refer to the parts quadratic in the first-order quantities as \emph{sources} and to the linear structure of the differential system as \emph{couplings}; note that, at first order, only the latter are present.
The vector and tensor modes of the metric do not {\em directly} induce B-polarisation \cite{beneke:2010a}, meaning that the coupling between the B-modes and the metric vanishes, $\,\mathcal{M}_{B,\lm}=0\,$. Similarly, at the linear level, Compton scattering does not directly generate B-polarisation from other polarisation patterns, and therefore $\,\Gamma_{(B,\lm)n'}\Delta_{n'}=0\,$.
It follows that the only non-vanishing coupling in the B-mode hierarchy is the free streaming, $\,\Sigma_{(B,\lm)n'}\,\Delta_{n'}\,$, while the only sources are the quadratic Liouville and scattering sources, $\,\mathcal{Q}^L_{B,\lm}\,$ and $\,\mathcal{Q}^C_{B,\lm}\,$, which read\footnote{The quadratic sources of the Boltzmann equation are a convolution integral over two dummy wavemodes, \kone and \ktwo. In the following, for brevity, we shall report only the kernels of the convolution and assume that the first term in a product is assigned $\kone$ and the second $\ktwo$, \eg $\Phi\,\Delta_{E,\L m_1}=\Phi(\kone)\,\Delta_{E,\L m_1}(\ktwo)\,$.}
\begin{align}
  \label{eq:boltzmann_pure_bmodes}
  &\begin{aligned}
    \Sigma_{(B,\lm)n'}\,\Delta_{n'} \;=\;
      \Delta_{B,\L+1\,m}\,D^{+,\L}_{m\,m}\;
      -\;\Delta_{B,\L-1\,m}\,D^{-,\L}_{m\,m}
      -\;\Delta_{E,\L\,m}\,D^{0,\L}_{m\,m}\;,
  \end{aligned}
  \\[0.44cm]
  \label{eq:boltzmann_quad_liouville_bmodes}
  &\begin{aligned}
    \mathcal{Q}^L_{B,\lm} \;=\;
    i\;\Bigl\{\;
    k_2^{[m_2]}\left(\Psi+\Phi\right)\;
    +\;4\;k_1^{[m_2]}\,\Psi\;
    -\;k_1^{[m_2]}\left(\,\Psi\,+\,\Phi\,\right)
    \;\Bigr\}\;
    D^{0,\ell}_{m_1\,m}\;\Delta_{E,\L m_1} \,
  \end{aligned}
  \msk
  \label{eq:boltzmann_quad_collision_bmodes}
  &\begin{aligned}
    &\mathcal{Q}^C_{B,\lm} \;=\;
    \dot\kappa\;v_e^{[m_2]}
    \;\Bigl\{ \;
    \Delta_{E,\ell m_1}\;
    -\;\delta_{\ell2}\;\frac{\sqrt{6}}{5}\,
    (\Delta_{I,2m_1}-\sqrt{6}\,\Delta_{E,2m_1}) \; \Bigr\} 
    \:D^{0,\ell}_{m_1\,m} \;,
  \end{aligned}
\end{align}
where we have followed the notation in Ref.~\cite{beneke:2010a, beneke:2011a}.
In particular, the indices $\,m_1\,$ and $\,m_2=m-m_1\,$ are implicitly summed
over $m_1=-1,0,+1$: scalar, vector and tensor modes are mixed at second
order.
The coupling coefficients $D$ are shorthand for the multipole decompositions of
$\,n^if\,$ and of $\,(\delta_{ij}-n_in_j)\,\partial f/\partial n^j\,$; we report their explicit form in Appendix~\ref{app:coupling}.

The free-streaming matrix $\Sigma_{(B,lm)n'}$
in \eref{eq:boltzmann_pure_bmodes} couples E and B modes, so that
the two types of polarisation source each other.
This coupling is effective in converting E to B-polarisation all the way from
recombination to today.
Therefore, the sources appearing in the hierarchy for the E-modes are as
efficient as the ones in \eref{eq:boltzmann_quad_liouville_bmodes} and
\eref{eq:boltzmann_quad_collision_bmodes} in generating the B-modes.
For this reason, we shall refer to the quadratic sources $\mathcal{Q}^C_{B,\lm}$,
$\mathcal{Q}^L_{B,\lm}$, $\mathcal{Q}^C_{E,\lm}$ and $\mathcal{Q}^L_{E,\lm}$ as
\emph{direct sources} for the B-modes. 


As in the first-order case, the intensity and E-polarisation quadrupoles are coupled via   $\Gamma_{(E,lm)n'}\Delta_{n'}$, whose expression can be found in \eref{eq:compact_collision_term_gammamatrix}.
This coupling is active only during recombination as $\Gamma_{nn'}$ multiplies the Compton scattering rate.
Hence, all non-scalar sources that affect the intensity will lead to the
generation of B-modes in an indirect way: during recombination they are converted into E-modes, which in turn couple to the B-modes by the efficient free-streaming mechanism encoded by $\Sigma_{nn'}$. For this reason, we shall call such sources \emph{indirect}.
The indirect sources include $\mathcal{Q}^C_{I,\lm}$ and $\mathcal{Q}^L_{I,\lm}$, but also the
metric sources $\mathcal{M}_{I,\lm}$.
Since all species are coupled to the metric, this means that any non-scalar source (\eg from the cold dark matter and neutrino sectors) will be indirectly coupled to B-mode polarisation if it is active before or during recombination. This is indeed the way B-polarisation is generated at first order if a background of primordial gravitational waves is present.

\section{Line-of-sight}
\label{sec:line_of_sight}

After recombination, photons stream so that at time $\eta$ higher multipoles with $\ell\approx k
(\eta-\eta_{\text{rec}})$ are excited, making a full numerical computation of
the photon brightness moments $\Delta_{\lm}$ increasingly expensive. In \SONG we instead compute the moments using the \los integration
\begin{equation}
	\label{eq:los_integral}
	\Delta_{n}(\eta_0) 
	= \int_0^{\eta_0}  d\eta \,e^{-\kappa(\eta)}
j_{nn'}(k(\eta_0-\eta))\,\mathcal{S}_{n'}(\eta) \;,
\end{equation}
which is an integral representation for the solution of the differential
equation \cite{seljak:1996a}
\begin{equation} \label{eq:Losdgl}
	\dot{\Delta}_{n} + k \, \Sigma_{nn'}\Delta_{n'} =
-|\dot{\kappa}|\Delta_{n} 
	+\mathcal{S}_n \;.
\end{equation}
The source function $\mathcal{S}_{n}$ contains both couplings and quadratic sources, and it is obtained by comparing equation (\ref{eq:Losdgl}) with the full Boltzmann equation (\ref{eq:boltzmann_compact}):
\begin{equation}
	\label{eq:line_of_sight_sources}
	\mathcal{S}_n =-\mathcal{M}_{n} -\mathcal{Q}^L_{n}+
|\dot{\kappa}|\left(\Gamma_{nn'}\Delta_{n'}+\mathcal{Q}^C_{n}\right) \;.
\end{equation}

The excitation of higher multipoles through streaming and the mixing of E- and B-modes is encoded in the projection functions $j_{nn'}$, which are linear combinations of spherical Bessel functions \cite{beneke:2011a}.
Because the free-streaming coupling does not mix intensity with polarisation (as shown in \eref{eq:boltzmann_pure_bmodes}) the projection function $j_{(B,\lm)n'}$ vanishes for $n'=(I,\L'm')$. As a result, to compute $\,\Delta_{B,\lm}(\eta_0)\,$ one only needs to integrate the source functions for the E and B-modes, that is, $\,\mathcal{S}_{E,\lmp}\,$ and $\,\mathcal{S}_{B,\lmp}\,$.

The quadratic part of the line-of-sight sources for polarisation is given by $\,-\mathcal{Q}^L_{B,\lm}+|\dot{\kappa}|\mathcal{Q}^C_{B,\lm}$ and $ -\mathcal{Q}^L_{E,\lm}+|\dot{\kappa}|\mathcal{Q}^C_{E,\lm}\,$, which we identified in the previous section as direct sources. These depend only on linear perturbations squared and are therefore easy to compute.
The metric terms in $\mathcal{M}_{n}$ vanish for both polarisation types, while $\,|\dot{\kappa}|\Gamma_{(B,\lm)n'}=0\,$ only for the B-modes. Therefore, the only other contribution to $\,\Delta_{B,\lm}(\eta_0)\,$ comes from  $|\dot{\kappa}|\Gamma_{(E,\lm)n'}\Delta_{n'}\,$, which involves the second-order photon perturbation $\Delta_{n'}$.
This term encodes the indirect generation of B-modes by sources that affect the intensity or the E-modes, that is, the indirect sources.
For B-polarisation, these are only relevant before recombination as they depend on Compton scattering. To obtain them one has to solve the second-order Boltzmann-Einstein differential system until the time of recombination; for this purpose, we use the second-order Boltzmann code \SONG \cite{pettinari:2013a}.

We split our computation into two steps. First we solve the full differential equation system (\ref{eq:boltzmann_compact}) including all second-order sources (metric, collision and Liouville) up to the end of recombination. This is possible because during the tight-coupling regime all the multipoles with $\ell>2$ are suppressed and, towards the end of recombination, they are only slowly generated. Thus we can cut the hierarchy at relatively small $\ell\approx 15$ (for details on this step we refer to Ref.~\cite{pettinari:2013a}).
The resulting transfer functions are used to construct the indirect sources, $\,|\dot{\kappa}|\Gamma_{(E,\lm)n'}\Delta_{n'}\,$, which are added to the direct ones to obtain the line-of-sight source functions $S_{n'}\,$. Finally, these are numerically convolved with the projection functions $\,j_{nn'}\,$ in \eref{eq:los_integral} to obtain the B-mode polarisation today.

The lensing and time-delay terms in $\mathcal{Q}^L_n$ are impractical to compute numerically in the line-of-sight formalism, as they involve photon moments that are active all the way to today. A full treatment of these terms is beyond the scope of this work and non-perturbative computations for these effects have already been performed. 
For these reasons, in this work we neglect the lensing and time-delay terms and refer to the non-perturbative results for lensing \cite{zaldarriaga:1998c, seljak:2004a, lewis:2006a, smith:2009a} and time-delay \cite{hu:2000a}.
We further refer the reader to Huang \& Vernizzi (2013) \cite{huang:2013b} for a second-order treatment of such terms for the photon intensity.

\subsection{Treating the redshift contribution}
\label{sec:delta_tilde}

For the intensity, it was shown by Huang \& Vernizzi \cite{huang:2013a} that the
quadratic redshift contribution, that is, the second term in \eref{eq:QL}, can
be included in the computation by a transformation of variables (see also Ref.~\cite{pettinari:2013a} or, for an alternative treatment of the redshift terms, Ref.~\cite{su:2012a}).
Here, we generalise this approach to polarised radiation and consider the
following transformation:
\begin{equation}
	\label{eq:delta_tilde_transformation}
 	\widetilde{\Delta}_{ab} \;\equiv\; [\,\ln(1 + \Delta)\,]_{ab} \;. 
\end{equation}
The logarithm of matrices is defined by the Mercator series; up to second order we obtain:
\begin{equation}
	\label{eq:delta_tilde_transformation_2nd}
	\widetilde{\Delta}_{ab} \;=\; \Delta_{ab} \;-\; \frac{\Delta_{ac}\,\Delta_{cb}}{2}\;.
\end{equation}
Expanding the above expression in multipole space, and neglecting the contribution from the first-order B-modes, gives rise to a non-trivial mixing between the $I,E,B$ modes:
\begin{align}
  \label{eq:delta_tilde_transformation_multipole_space_B}
  \widetilde{\Delta}_{B,\lm} \;&=\; \Delta_{B,\lm} - i^{L-1}
  \left(\begin{array}{cc|c}\L'&\L''&\L \\ m' & m'' & m\end{array}\right)
  \left(\begin{array}{cc|c}\L'&\L''&\L \\ 0 & 2 &
  2\end{array}\right)\Delta_{I,\L'm'}\Delta_{E,\L''m''} & \text{ if } L \text{
  odd}\;,
  \msk
  \label{eq:delta_tilde_transformation_multipole_space_E}
  \widetilde{\Delta}_{E,\lm} \;&=\; \Delta_{E,\lm} - i^{L}
  \left(\begin{array}{cc|c}\L'&\L''&\L \\ m' & m'' & m\end{array}\right)
  \left(\begin{array}{cc|c}\L'&\L''&\L \\ 0 & 2 &
  2\end{array}\right)\Delta_{I,\L'm'}\Delta_{E,\L''m''} & \text{ if } L \text{
  even}\;,
\end{align}
and $ \widetilde{\Delta}_{B,\lm} = \Delta_{B,\lm}$,  $\;\widetilde{\Delta}_{E,\lm} = \Delta_{E,\lm}\,$ otherwise with $L=\L-\L'-\L''$ and where
\begin{equation}
  \left(\begin{array}{cc|c}\L'&\L''&\L \\ m' & m'' & m\end{array}\right)
  \;=\; \avg{\L'm'\L''m''|\L m}
\end{equation}
are the Clebsch-Gordan coefficients.
A sum over $\ell',\ell'',m',m''$ is implicit, a convention that will be assumed also for the equations that follow.
The transformation for the unpolarised brightness reads
\begin{align}
  \label{eq:delta_tilde_transformation_multipole_space_I}
  \widetilde{\Delta}_{I,\lm} \;=\; \Delta_{I,\lm} \;&- \frac{1}{2}\, i^{L}\,
  \left(\begin{array}{cc|c}\L'&\L''&\L \\ m' & m'' & m\end{array}\right)
  \left(\begin{array}{cc|c}\L'&\L''&\L \\ 0 & 0 & 0\end{array}\right)
  \Delta_{I,\L'm'}\Delta_{I,\L''m''}
  \nmsk
  &- \frac{1}{2} \,i^{L}\,
  \left(\begin{array}{cc|c}\L'&\L''&\L \\ m' & m'' & m\end{array}\right)
  \left(\begin{array}{cc|c}\L'&\L''&\L \\ 2 & -2 & 0\end{array}\right)
  \Delta_{E,\L'm'}\Delta_{E,\L''m''} &\text{ if } L \text{ even} \;,
\end{align}
and $\widetilde{\Delta}_{I,\lm} = \Delta_{I,\lm}$ otherwise.
It can be shown that it is not necessary to include the $\Delta_{E,\L'm'}\Delta_{E,\L''m''}$ term in the second line in order to absorb the redshift term for the intensity; indeed, Refs. \cite{pettinari:2013a, huang:2013a} neglect it.

The time derivative of $\widetilde{\Delta}_{ab}$ up to second order is then given by
\begin{eqnarray}
	\nonumber
	\dot{\widetilde{\Delta}}_{ab} \;=\; \left[\dot{\Delta} -
\frac{\dot{\Delta}\Delta +\Delta\dot{\Delta}}{2} \right]_{ab}&\;=\;&- n_i \partial^i\widetilde{\Delta}_{ab} \,-\,
\mathcal{M}_{ab} \,+\,\left[\mathfrak{C}-\frac{\mathfrak{C}\,\Delta+\Delta\,\mathfrak{C}}{2}\right]_{ab} \\[0.25cm]
	&& -\,\mathcal{Q}^L_{ab} \,+\, 4 \, (n^i\partial_i \Psi
-\dot{\Phi})\Delta_{ab} \;,
  \label{eq:absorption}
\end{eqnarray}
where we have used the first-order Boltzmann equation
\begin{equation}
	\dot{\Delta}_{ab} \;=\; -n_i\partial^i \Delta_{ab} \,-\, 4 \, (n^i\partial_i \Psi
\,-\,\dot{\Phi})\,\delta_{ab} \,+\, \mathfrak{C}_{ab}  \,,
\end{equation}
to replace the quadratic terms including $\dot{\Delta}_{ab}$ and the second-order
one, \eref{eq:boltzmann_compact}, to replace $\dot{\Delta}_{ab}$.
The new contribution $\,4\, (n^i\partial_i \Psi -\dot{\Phi} )\Delta_{ab}\,$ exactly
cancels the redshift term in $\mathcal{Q}^L$ so that the second line of
\eref{eq:absorption} reduces to only the time-delay and lensing contributions.
In addition, the collision term $\mathfrak{C}_{ab}$ is replaced by
$[\mathfrak{C}-(\mathfrak{C}\Delta+\Delta\mathfrak{C})/2]_{ab} $.\footnote{
  Note that since $\tilde\Delta=\Delta$ to first order, we can replace 
  $\Delta$ by $\tilde\Delta$ in the non-linear terms, so that \eref{eq:absorption} becomes a closed system of equations for $\tilde\Delta$.}

The transformation works because the second-order source we are
eliminating is precisely the first-order $\Delta$ times part of the first-order source.
Its cost is that it makes the scattering term more complex and requires an additional transformation to relate $\widetilde{\Delta}_{ab}$ back to $\Delta_{ab}$.
Unfortunately, this still leaves other problematic terms in $\mathcal{Q}^L$, the
lensing and time-delay terms. 

\subsection{Temperature definition}
\label{sec:temperature_definition}


At second order it is not possible to unambiguously define a temperature for the
photon distribution function, as its blackbody spectrum is distorted by
collisions during recombination and reionisation \cite{dodelson:1995a, hu:1994a, hu:2000a, bartolo:2006a, pitrou:2010b, pettinari:2013a, renaux-petel:2013a}.
However, one can choose between a number of ``effective'' temperatures, each
corresponding to a different moment of the distribution function; in the limit
of a blackbody spectrum these all coincide.
In this paper, we adopt the so-called bolometric temperature $\Theta_{ab}$,
which is the temperature of a blackbody distribution with the same energy
density of the photons \cite{pitrou:2010b, pitrou:2014a}.
For polarised radiation, the brightness is related to the bolometric temperature by:
\begin{equation}
	\label{eq:bolometric_temperature}
	\delta_{ab} \,+\, \Delta_{ab} \;=\; (\delta\,+\,\Theta)_{ab}^4 \;,
\end{equation}
Note that this definition connects the bolometric temperature for the intensity, E-mode and B-mode polarisation to the brightness moments in a non-trivial way.
Expanding this relation to second order,
\begin{align}
  \label{eq:bolometric_temperature_2nd}
  \Delta_{ab} \;=\; 4\,\Theta_{ab} \;+\; 6\,\Theta_{ac}\,\Theta_{cb} \;,
\end{align}
we obtain the product $ \Theta_{ac} \, \Theta_{cb}$, which is expanded into multipole space similarly to \eref{eq:delta_tilde_transformation_multipole_space_B} and thus
contains quadratic couplings between the B-mode, the E-mode and intensity.
Note that $[\Theta^2]_{ab}$ contributes to the B-polarisation component of 
$\Delta_{ab}$ at second order even in the absence of B-polarisation of the temperature anisotropies at first order.
This effect is present regardless of the
choice of temperature definition; in particular, it exists even for a blackbody spectrum where all
choices are equivalent. 

\section{Power spectrum computation}
\label{sec:spectrum}

The angular power spectrum of the intrinsic B-polarisation, $C_\ell^{BB}$, is
defined as
\begin{align}
  \avg{a_{B,\ell m}\,a^*_{B,\ell' m'}} \;=\;
C_\ell^{BB}\,\delta_{\ell\ell'}\,\delta_{mm'} \;,
\end{align}
where the observed multipoles $a_{X,\lm}$, with $X=I,E,B$, are
conventionally related to the temperature perturbation $\Theta_{X,\lm}$ by
\footnote{The extra $\ell$-dependent factors in the integral counter the factors
included in the multipole decomposition of $\Theta_{\ell m}$ in order to
simplify the Boltzmann equations. Also note that the position $\vec{x_0}$ where the $a_{\ell m}$'s are computed is not important,
as it will cancel in the power spectrum after enforcing the statistical
homogeneity of the Universe.
}
\begin{equation}
  \label{eq:alm_definition}
  a_{X,\ell m}\;=\;
    \int \frac{\text{d}^3\vec{k}}{(2\pi)^3}\,(-i)^{\L}\sqrt{\frac{4\pi}{2\ell+1}}\;
    \Theta_{X,\ell m}(\vec{k})\;e^{i\vec{k}\cdot\vec{x_0}} \;.
\end{equation}
For purely scalar initial conditions, no B-polarisation is generated at first order ($a_{B,\ell m}^{(1)}=0$) and, as a result, $C_\ell^{BB}$ is fourth-order in the primordial perturbations. For the same reason, the contribution to $C_\ell^{BB}$ from the three-point function is at least fifth order and we neglect it.

The correction needed to absorb the redshift term from the Boltzmann equation
Eq.~(\ref{eq:delta_tilde_transformation_2nd}) and the quadratic term in the relation Eq.~(\ref{eq:bolometric_temperature_2nd}) between brightness and 
the bolometric temperature have the same structure. After enforcing $\Delta_{ab}=4\,\Theta_{ab}$ up to first order, they
partially cancel to yield an expression for the second-order temperature perturbation:\footnote{In \BF the quadratic term in the relation (\ref{eq:bolometric_temperature_2nd}) between brightness and bolometric temperature was mistakenly omitted and the relation  $\Theta_{ab}^{(2)} = \frac{1}{4}\,\Delta_{ab}^{(2)}$ was used to compute 
the angular power spectrum. Given the above-mentioned cancellation 
with the redshift contribution, which was neglected in \BF, 
their numerical result is fortuitously close to that where both effects are included.}
\begin{align}
  \label{eq:temperature_perturbation_redshift_plus_temperature}
  \Theta^{(2)}_{ab} \;=\; \frac{1}{4}\,\widetilde{\Delta}^{(2)}_{ab}
  \;+\; \frac{1}{2}\,\Theta_{ac}^{(1)}\,\Theta_{cb}^{(1)} \;,
\end{align}
where the partial cancellation is encoded by the factor $\nicefrac{1}{2}=2-\nicefrac{3}{2}$, and we have reinstated the suffix denoting the perturbative order to avoid confusion.
We recall that $\widetilde{\Delta}_{ab}$ is the transformed distribution
function that was introduced in \sref{sec:delta_tilde} to treat the redshift
contribution of the Boltzmann equation.
The helicity indices $a,b,c$ can be converted to polarisation multipoles using
\eref{eq:multipole_decomposition}. For the B-polarisation, this results in
\begin{equation}
  \label{eq:temperature_perturbation_redshift_plus_temperature_multipole}
  \Theta_{B,\lm}^{(2)} \;=\; \frac{1}{4}\,\widetilde{\Delta}_{B,\lm}^{(2)} \,-\,
i^{\ell-\ell'-\ell''+1}
    \left(\begin{array}{cc|c}\ell'&\ell''&\ell \\ m' & m'' & m\end{array}\right)
    \left(\begin{array}{cc|c}\ell'&\ell''&\ell \\ 0 & 2 & 2\end{array}\right)
    \;\Theta_{I,\ell'm'}^{(1)}\;\Theta_{E,\ell''m''}^{(1)} \;
\end{equation}
for $\L-\L'-\L''$ odd, while for even $\L-\L'-\L''$ the non-linear term is absent.
(We recall that, here and in the following, a sum is implicit over the primed indices, as well as a convolution product over the Fourier wave-vector.)
The computation of the angular power spectrum $C_\ell^{BB}$ is straightforward as our assumption is that there are no first-order B-modes. The leading term is quadratic in $\Theta_{B,\lm}^{(2)}$ and will therefore contain three contributions:
\begin{equation}
  \label{eq:power_spectrum_three_terms}
  C_\ell^{BB} \;=\; \frac{1}{16}\,C_{\ell}^{\widetilde{B}\widetilde{B}} 
  \;-\; \frac{1}{2}\,C_{\ell}^{\widetilde{B}(IE)} \;+\; C_{\ell}^{(IE)(IE)} \;.
\end{equation}

The first term in \eref{eq:power_spectrum_three_terms} is the power spectrum of
the transformed B-modes $\widetilde{\Delta}^{(2)}_{B,\lm}$:
\begin{equation}
   C_{\ell}^{\widetilde{B}\widetilde{B}} \;=\;
   \int \frac{d^3\vec{k}}{(2\pi)^3}\frac{d^3\vec{k}'}{(2\pi)^3} \frac{4\pi}{2\ell+1} 
   \avg{\widetilde{\Delta}_{B,\lm}^{(2)}(\vec{k})\:\widetilde{\Delta}^{(2)*}_{B,\lm}(\vec{k}')} \;,
  \label{eq:cl_bb}
\end{equation}
with no summation over $m$.\footnote{The power spectrum $C_{\ell}^{\widetilde{B}\widetilde{B}}$ does not depend on the value of the azimuthal mode $m$ because there is no preferred direction. In the following we shall choose to rotate the coordinate system so to align the symmetry axis of the spherical harmonics to $\vec{k}$; this does not change the previous statement, but $C_{\ell}^{\widetilde{B}\widetilde{B}}$ will then be expressed as a sum over the different $m$ modes. We shall call the $m=\pm1$ and $m=\pm2$ contributions as vector and tensor modes, respectively.}
To compute this power spectrum we rotate the coordinate system such that the polar axis of the spherical harmonics, $\,\vec{\hat z}\,$, is aligned with the Fourier mode $\vec{k}\,$; the perturbation
in the new coordinate system is given by a Wigner rotation, which in turn is expressed using spin-weighted spherical harmonics \cite{hu:1997b, pitrou:2010a, beneke:2011a}:
\begin{align}
  \pert{\widetilde{\Delta}}{2}_{B,\lm}(\vec{k}) \quad&\longrightarrow\quad
  \sqrt{\frac{4\,\pi}{2\,\ell+1}}\; \sum\limits_{m'=-\ell}^{\ell}
  Y_{\lm}^{-m'}(\vec{\hat{k}})\;\,\pert{\widetilde{\Delta}}{2}_{B,\ell m'}(k\,\vec{\hat z}) \;.
  \label{eq:wigner_rotation}
\end{align}
The statistical isotropy of the Universe ensures that all the relevant physical information is contained in $\,\pert{\widetilde{\Delta}}{2}_{B,\ell m}(k\,\vec{\hat z})\,$, which hereafter we shall simply denote as $\,\pert{\widetilde{\Delta}}{2}_{B,\ell m}(k)$.
We also express the second-order perturbation $\widetilde{\Delta}^{(2)}_{B,\lm}(\vec{k})$ in terms of the \emph{second-order transfer function} $\widetilde{T}^{(2)}_{B,\lm}(\vec{k},\vec{k}_1,\vec{k}_2)$ and the primordial perturbation $\Phi$:
\begin{equation}
\widetilde{\Delta}_{B,\lm}^{(2)}(\vec{k}) \;=\; \int \frac{d^3 \vec{k}_1}{(2\pi)^3}\; \widetilde{T}^{(2)}_{B,\lm}(\vec{k},\vec{k}_1,\vec{k}_2) \, \Phi(\vec{k}_1) \, \Phi(\vec{k}_2)\;,
\label{eq:transfer_function}
\end{equation}
with $\vec{k}_2= \vec{k}-\vec{k}_1$. Finally, we use Wick's theorem to express the expectation values of the four primordial perturbations resulting from substituting \eref{eq:transfer_function} into \eref{eq:cl_bb} as a product of two power spectra, $\,P_\Phi\,$.
%
%
We thus obtain an expression that only contains quantities that can be computed by \SONG:
\begin{eqnarray}
  C_{\L}^{\widetilde{B}\widetilde{B}} &\;=\;& \frac{2}{\pi} \,  \frac{1}{(2 \L+1)^2} \;\sum \limits_{m \neq 0} \;
  \int dk\,k^2 \int \frac{d^3 \vec{k}_1}{(2\pi)^3} \; P_{\Phi}(k_1)\, P_{\Phi}(k_2) \nmsk
  &&  \left(\;
  \widetilde{T}^{(2)}_{B,lm}(k,\vec{k}_1,\vec{k}_2)\:\widetilde{T}^{(2)*}_{B,lm}(k,\vec{k}_1,\vec{k}_2)
  \;+\;
  \widetilde{T}^{(2)}_{B,lm}(k,\vec{k}_1,\vec{k}_2)\:\widetilde{T}^{(2)*}_{B,lm}(k,\vec{k}_2,\vec{k}_1)
  \;\right)\,.
\label{eq:CBBsummed}
\end{eqnarray}
This formula does not involve scalar contributions ($m=0$), as the scalar modes do not generate B-mode polarisation.
In principle, the sum over $m$ goes as far as $m=\pm\ell\,$ but, in practice, there are two reasons why the dominant contribution is expected to come from the vector ($m=\pm1$) and tensor ($m=\pm2$) parts.
First, the $|m|>2$ contributions only arise from $\ell>2$ multipoles which are tight-coupling suppressed during recombination.
Secondly, being $\,Y_{\lm}(\theta,\phi)\,$ proportional to $\,\sin^m\theta\,$, the $|m|>0$ sources are increasingly suppressed for ``squeezed'' configurations, where either one of the $\kone$ or $\ktwo$ wavevectors is aligned with the polar axis of the spherical harmonics, $\vec{k}$.
This suppression diminishes as $\kone$ and $\ktwo$ become more orthogonal with respect to $\vec{k}$.
Because the three wavevectors form a triangle, $\vec{k}=\kone+\ktwo$, these large-angle configurations correspond to increasingly large values of $k_1$ and $k_2$, which, in turn, are suppressed by the $P_{\Phi}(k_1)\,$ and $\,P_{\Phi}(k_2)$ terms in the power spectrum formula.
We have verified the above arguments numerically by running \SONG with sources up to $|m|=4$, and found that the $m=\pm3$ and $m=\pm4$ sources indeed give a negligible total contribution of about $1\%$ to $C_{\L}^{\widetilde{B}\widetilde{B}}$.

The second term in $C_\ell^{BB}$ is the mixed contribution $\,C_{\ell}^{\widetilde{B}(IE)}\,$, given by the statistical average of the product of $\widetilde{\Delta}^{(2)}_{B,\lm}$ with the quadratic part of \eref{eq:temperature_perturbation_redshift_plus_temperature_multipole}; it takes the form of a bispectrum that is folded with two Clebsch-Gordan coefficients:
\begin{eqnarray} \nonumber
  C_{\ell}^{\widetilde{B}(IE)} &=&
    \frac{4\pi}{2\ell+1}\;i^{\ell-\ell'-\ell''+1}\,
    \left(\begin{array}{cc|c}\ell'&\ell''&\ell \\ m' & m'' & m\end{array}\right)
    \left(\begin{array}{cc|c}\ell'&\ell''&\ell \\ 0 & 2 & 2\end{array}\right)
    \int \frac{d^3\vec{k}}{(2\pi)^3}\frac{d^3\vec{k}'}{(2\pi)^3}
    \frac{d^3\hat{\vec{k}}}{(2\pi)^3}\frac{d^3\hat{\vec{k}}'}{(2\pi)^3}\msk
    &&(2\pi)^3\delta(\vec{k}-\hat{\vec{k}}-\hat{\vec{k}}')
    \avg{\Theta_{I,\ell'{m}'}^{(1)}(\hat{\vec{k}})\:\Theta_{E,\ell''{m}''}^{(1)}
    (\hat{\vec{k}}')\: \widetilde{\Delta}^{(2)*}_{B,\lm}(\vec{k}')} \;.
  \label{eq:cl_bie}
\end{eqnarray}
After introducing the transfer functions \eref{eq:transfer_function} and rotating the coordinate system so that the z-axis is aligned with $\vec{k}'$, we are able to express $\,C_{\ell}^{\widetilde{B}(IE)}\,$ in terms of the quantities computed by \SONG:
\begin{eqnarray} \nonumber
  C_{\ell}^{\widetilde{B}(IE)} &\,=\, &
    i^{\ell-\ell'-\ell''+1}\,\frac{4\pi}{(2\ell+1)}\,\frac{2}{\pi}\,
    \sum \limits_{m \neq 0} \left(\begin{array}{cc|c}\ell'&\ell''&\ell \\ m' & m'' & m\end{array}\right)
    \left(\begin{array}{cc|c}\ell'&\ell''&\ell \\ 0 & 2 & 2\end{array}\right) \\[0.20cm]
    &&\int dk k^2\int
    \frac{d^3\vec{k}_1}{(2\pi)^3}  P_\Phi(k_1) P_\Phi(k_2)  \; \frac{ Y_{\ell'm'}(\vec{k}_1) }{\sqrt{2\ell'+1}} 
   \;\frac{Y_{\ell''m''}(\vec{k}_2)}{\sqrt{2\ell''+1}}\; \nmsk
   && \frac{1}{4}\,T^{(1)}_{I,\ell'0}(k_1)  \;\frac{1}{4}\,T^{(1)}_{E,\ell''0}(k_2) \;
   \left(\;\tilde{T}^{(2)*}_{B,\ell m}(k,\vec{k}_1,\vec{k}_2) \,+\,
   \tilde{T}^{(2)*}_{B,\ell m}(k,\vec{k}_2,\vec{k}_1)\; \right) \;,
\end{eqnarray}
where we have expressed the first-order transfer functions $T^{(1)}_{\ell' m'}(\vec{k}')$ by their scalar component $ \sqrt{4\pi / (2\ell'+1)} \, Y_{\ell' m'}(\vec{k}')\, T^{(1)}_{\ell' 0}(k') $. 
The integration is numerically challenging because it is highly oscillatory in all directions. 
The second-order perturbation $\,\widetilde{T}_{B,\lm}^{(2)}\,$ depends on an additional Fourier wavevector, so that the integrations in $C_{\ell}^{\widetilde{B}(IE)}$ cannot be disentangled and computed separately.
To perform the integration we use the bispectrum module of \SONG, which will be
publicly released within the \emph{CLASS} code \cite{lesgourgues:2011a,
blas:2011a} in 2014 \cite{pettinari:2013c}.


The final term appearing in the intrinsic power spectrum comes from the square
of the quadratic part of
\eref{eq:temperature_perturbation_redshift_plus_temperature_multipole}.
This contribution depends only on the first-order perturbations and can be
reduced to the product of two linear power spectra:
\begin{eqnarray} 
  C_{\ell}^{(IE)(IE)} &\,=\,&
\frac{4\pi}{2\ell+1}i^{-\ell'-\ell''+\hat{\ell}'+\hat{\ell}''}
  \left(\begin{array}{cc|c}\ell'&\ell''&\ell \\ m' & m'' & m\end{array}\right)
  \left(\begin{array}{cc|c}\ell'&\ell''&\ell \\ 0 & 2 & 2\end{array}\right)
  \left(\begin{array}{cc|c}\hat{\ell}'&\hat{\ell}''&\ell \\ \hat{m}' & \hat{m}''
& m\end{array}\right)
  \left(\begin{array}{cc|c}\hat{\ell}'&\hat{\ell}''&\ell \\ 0 & 2 &
2\end{array}\right) \msk \nonumber
  &&\int
\frac{d^3\vec{k}}{(2\pi)^3}\frac{d^3\vec{k}'}{(2\pi)^3}\frac{d^3\hat{\vec{k}}}{
(2\pi)^3} 
\frac{d^3\hat{\vec{k}}'}{(2\pi)^3}\frac{d^3\widetilde{\vec{k}}}{(2\pi)^3}\frac{
d^3\widetilde{\vec{k}}'}{(2\pi)^3}
  \;(2\pi)^3\,\delta(\vec{k}-\hat{\vec{k}}-\hat{\vec{k}}')
  \;(2\pi)^3\,\delta(\vec{k}'-\widetilde{\vec{k}}-\widetilde{\vec{k}}') \msk
\nonumber 
  && \left<
\Theta^{(1)}_{I,\ell'm'}(\hat{\vec{k}})\:\Theta^{(1)}_{E,\ell''m''}(\hat{\vec{k}
}')\:
  \Theta^{(1)*}_{I,\hat{\ell}'\hat{m}'}(\widetilde{\vec{k}})\:
  \Theta^{(1)*}_{E,\hat{\ell}''\hat{m}''}(\widetilde{\vec{k}}')\right> \msk
\nonumber
  &&\hspace{-45pt}\;=\; \frac{1}{4\pi}\,\frac{(2\ell'+1)(2\ell''+1)}{2\ell+1}\;\left[\;
  \left(\begin{array}{cc|c}\ell'&\ell''&\ell \\ 0 & 2 &
2\end{array}\right)^2C_{\ell'}^{II}C_{\ell''}^{EE}+ 
  \left(\begin{array}{cc|c}\ell'&\ell''&\ell \\ 0 & 2 & 2\end{array}\right)
  \left(\begin{array}{cc|c}\ell'&\ell''&\ell \\ 2 & 0 &
2\end{array}\right)C_{\ell'}^{IE}C_{\ell''}^{EI}
  \;\right] \;.
  \label{eq:cl_ieie}
\end{eqnarray}
The three different contributions to the B-polarisation are all of comparable order; each is of fourth order in the perturbations and contains two polarised quantities that are suppressed compared to the unpolarised ones.


\section{Results}
\label{sec:results}

We compute the angular power spectrum of the intrinsic B-polarisation by first solving the Boltzmann equation \eref{eq:absorption} and then obtaining $C_\ell^{BB}$ via \eref{eq:power_spectrum_three_terms}.
In order to accomplish this task, we have updated \SONG with the Boltzmann treatment of polarised radiation at the level of the differential system and of the line-of-sight integration.
We have also implemented functions to compute polarised bispectra in an efficient way, as these are necessary to compute the $\,C_\ell^{B(IE)}\,$ contribution in \eref{eq:cl_bie}.
We obtain an independent check of our results by comparing \SONG's polarised transfer functions against a second numerical code based on Green functions. This code was used by \BF to compute the collisional intrinsic B-modes, and we have updated it since to include the metric and Liouville terms at recombination and the redshift terms along the line-of-sight.
Finally, we use adiabatic initial conditions with a vanishing tensor-to-scalar ratio, in agreement with \BF. Note that this implies that the initial conditions for the second-order vector and tensor perturbations vanish outside the Hubble horizon \cite{pitrou:2010a}.

\begin{figure}[p]
  \centering
    \includegraphics[height=0.38\textheight]{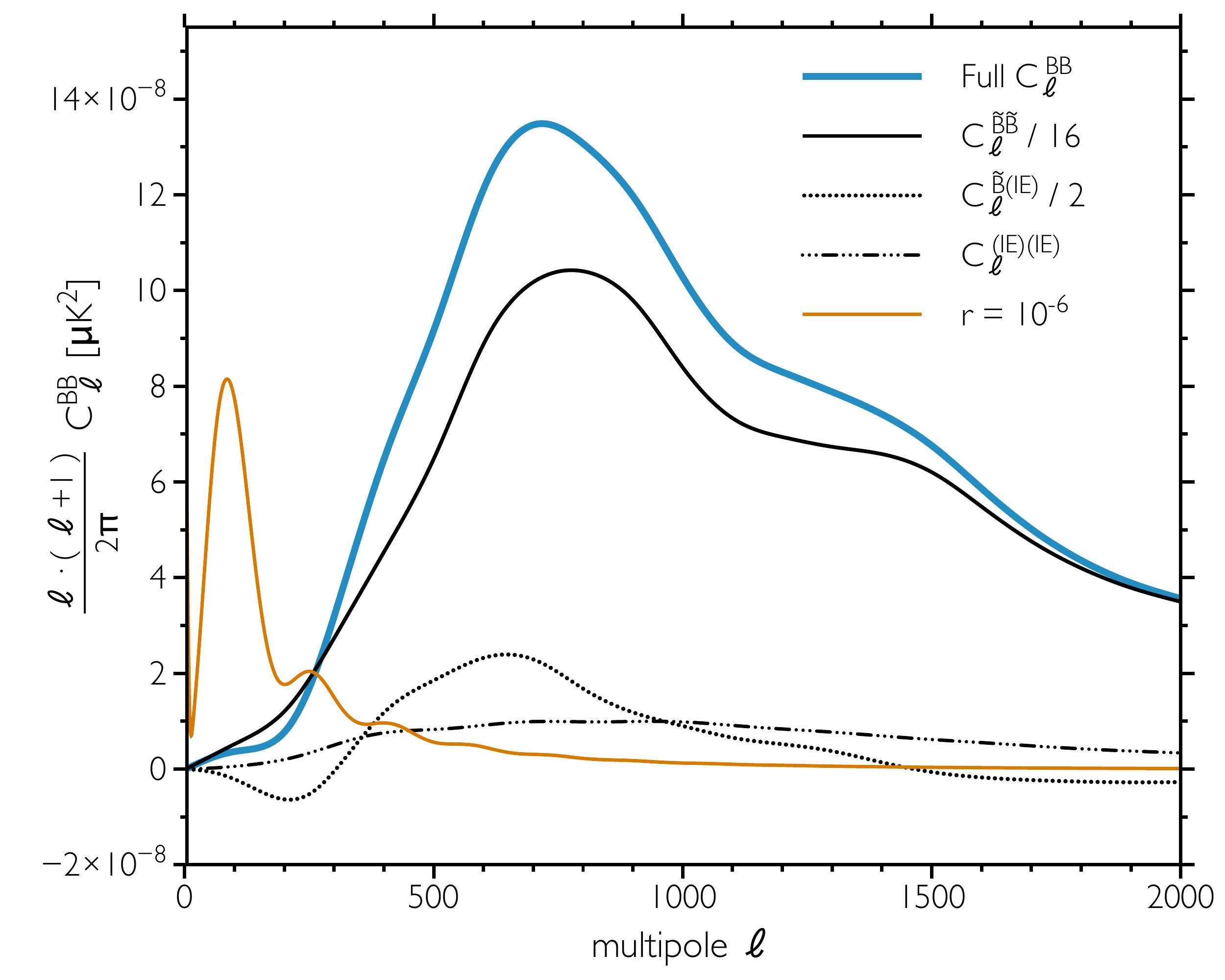}\\
    \includegraphics[height=0.38\textheight]{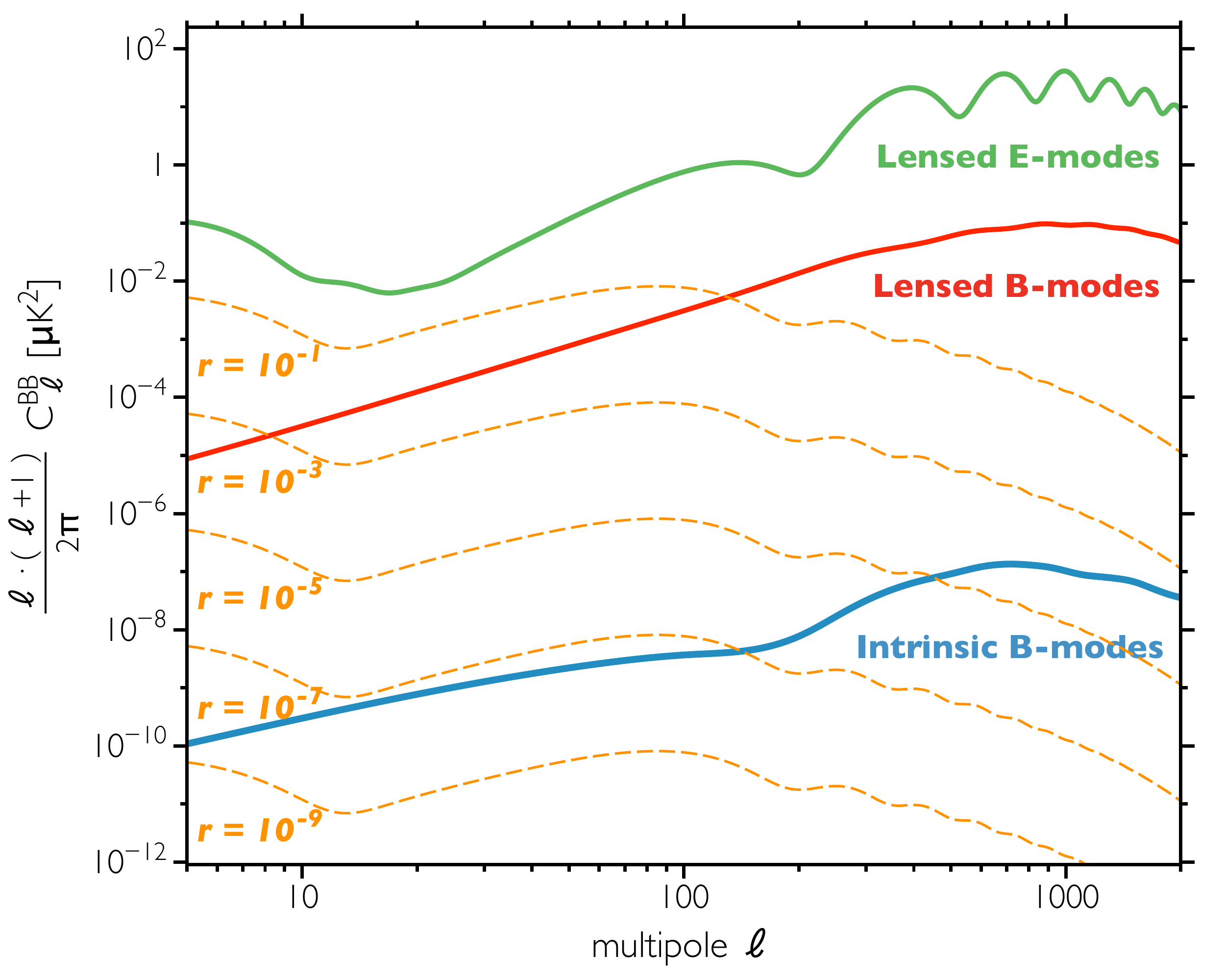}
  \caption{Angular power spectrum $C_\ell^{BB}$ of the intrinsic B-polarisation in the CMB. In this and in the following figures, we show the $C_\ell$'s multiplied by a factor $T_\text{cmb}^{\,2} = (\unit[\sci{2.7255}{6}]{\mu K})^2$ with respect to the equations in the text.
  \emph{Upper panel:} Breakdown of the intrinsic signal (solid blue curve) into its components, according to \eref{eq:power_spectrum_three_terms}. The $\widetilde{B}\widetilde{B}$ contribution (solid) dominates over the quadratic contributions $\widetilde{B}(IE)$ (dotted) and $(IE)(IE)$ (dot-dashed), due to the partial cancellation between the redshift and temperature corrections~\eref{eq:bolometric_temperature_2nd}, as explained in the text.
  \emph{Lower panel:} The intrinsic signal (solid blue curve) is comparable to that from primordial gravitational waves (dashed orange curves) with $r\approx10^{-7}$ at $\ell=100$ and $r\approx\sci{5}{-5}$ at $\ell\approx700$. It is always subdominant with respect to the lensing-induced B-modes (red curve) and the E-modes (green curve).
  }
  \label{fig:intrinsic_b_modes}
\end{figure}

In the upper panel of \fref{fig:intrinsic_b_modes} we show the full $C_\ell^{BB}$ together with a breakdown between its contributions according to \eref{eq:power_spectrum_three_terms}. The dominant component is given by $\widetilde{B}\widetilde{B}$, which arises from sources active at recombination. The other contributions, $\widetilde{B}(IE)$ and $(IE)(IE)$, arise from the bolometric-temperature correction \eref{eq:bolometric_temperature_2nd} and the redshift term \eref{eq:delta_tilde_transformation_2nd}. Their smallness with respect to $\widetilde{B}\widetilde{B}$ is due to the partial cancellation between the two effects, as explained after \eref{eq:temperature_perturbation_redshift_plus_temperature}. If we neglect the redshift contribution along the line-of-sight, the $C_\ell$ from $\widetilde{B}(IE)$ is three times larger and negative and the contribution from $(IE)(IE)$ is nine times larger and, as a result, the total $BB$ spectrum changes significantly in shape.

\clearpage

In the lower panel of \fref{fig:intrinsic_b_modes} we show a logarithmic comparison of the intrinsic B-modes with polarised signals from other sources. At its peak at $\ell=700$, the intrinsic signal is comparable to a primordial signal with tensor-to-scalar ratio of $r\approx\sci{5}{-5}$, while it corresponds to $r\approx10^{-7}$ at $\ell=100$, where the primordial signal peaks. The intrinsic spectrum is five to six orders of magnitude smaller than the the lensing-induced B-modes, which in turn is roughly $500$ times smaller than the E-modes spectrum at $\ell=700$. The B-polarisation signal from the time-delay effect, computed in Ref.~\cite{hu:2001a} and not shown in the figure, also peaks at $\ell\approx700$ and is approximately $80$ times larger than the intrinsic B-modes.

\begin{figure}[t]
  \centering
    \includegraphics[width=1\linewidth]{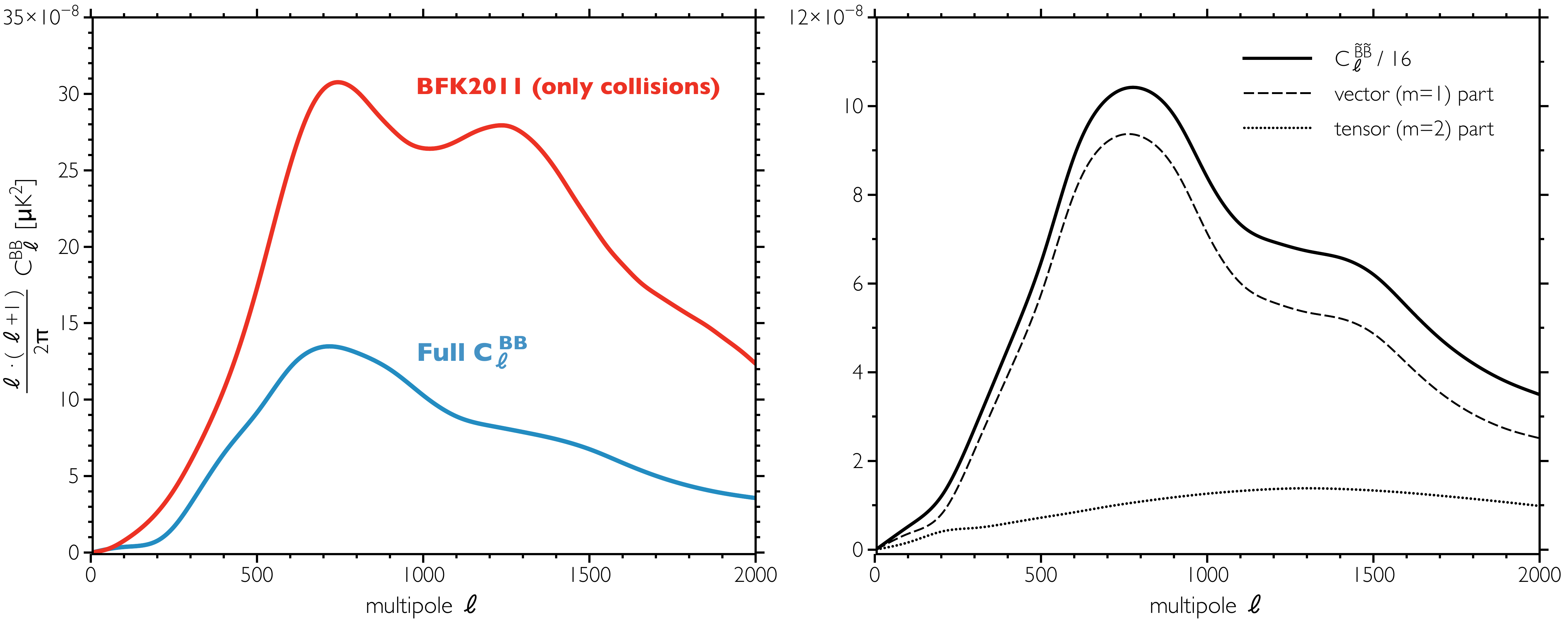}
  \caption{
  \emph{Left panel:} Comparison of the full intrinsic B-modes (solid blue curve) with a calculation including only collision effects and neglecting both the redshift and bolometric temperature corrections (solid red curve). The latter curve is consistent with what is computed by \BF.
  \emph{Right panel:} The power spectrum $C_\ell^{\tilde{B}\tilde{B}}$ (solid black curve) split into vector (dashed black curve) and tensor (dotted black curve) contributions. 
  }
  \label{fig:compare_b_modes}
\end{figure}

In the left panel of \fref{fig:compare_b_modes} we show the full intrinsic B-modes compared to the approximation used in \BF, considering only collision sources and neglecting both the redshift term and the temperature correction. In this case we find that the signal is enhanced by a factor of two. This is due to the modification of the collision term due to the absorption of the redshift terms, \eref{eq:absorption}. The extra collision term $-\,(\mathfrak{C}\Delta + \Delta \mathfrak{C})/2\,$ partially cancels with the quadratic sources in $\mathfrak{C}\,$. It should be noted that this is not the only effect of the redshift term, as the variable transformation needs to be reversed using the power spectra of $\widetilde{B}(IE)$ and $(IE)(IE)$. However, as discussed earlier these partially cancel with the temperature corrections, and together do not change the power spectrum significantly.

The right panel of \fref{fig:compare_b_modes} shows a split of $C_\L^{\tilde{B}\tilde{B}}$ into its vector and tensor contributions ($m=\pm1$ and $\pm2$ respectively in~\eref{eq:CBBsummed}). The tensor contribution is relatively smooth, while the vector contribution is similar in shape to that induced from collision sources in \BF. We conclude that the features are generated from the collision sources and are slightly distorted from the correlation with the other recombination effects.


\subsection{Comparison with previous results}
\label{sec:comparison_with_literature}

In the recent years several authors have treated parts of the full second-order computation. Here, we compare our results to their work.

\paragraph{Collision term} \BF quantified the generation of B-modes from the second-order Compton scattering during recombination. In order to reproduce their result, we restrict \SONG to include only the collisional sources, and to ignore the temperature and redshift corrections.
The resulting $C_\ell$ spectrum is shown in red in the left panel of \fref{fig:compare_b_modes}.
When using the same cosmology as in \BF, our result agrees with theirs within the statistical error bars resulting from the Monte-Carlo integration used in \BF. (\SONG does not have such statistical errors because it performs the integration using a different technique involving splines optimised for Bessel functions.) 

\begin{figure}[t]
	\centering
		\includegraphics[width=1\linewidth]{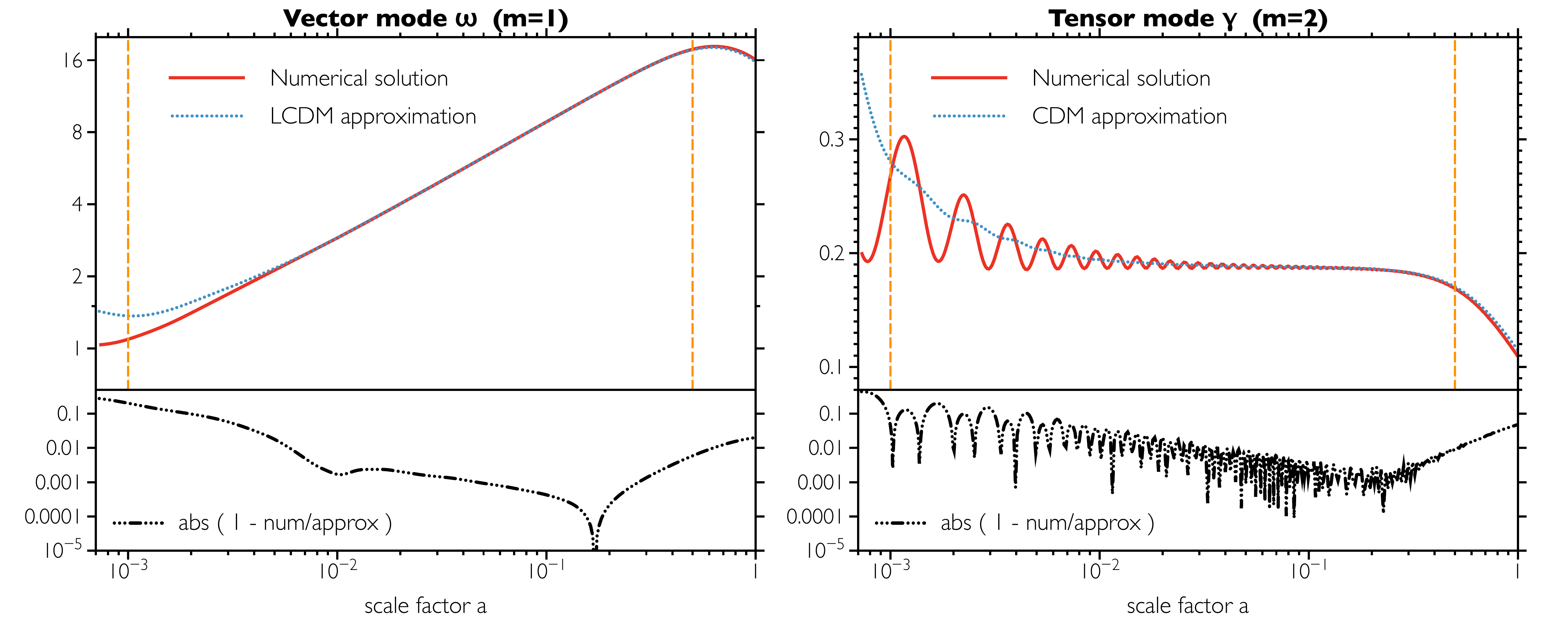}
	\caption{Evolution of the vector (left panel) and tensor (right panel) modes of the metric as obtained numerically by \SONG (continuous red curves) and approximately using the formulas in \MHM (dotted blue curves), for the Fourier mode $\,k_1=\unit[0.029]{Mpc^{-1}}$, $\,k_2=\unit[0.007]{Mpc^{-1}}$ and $\,k_3=\unit[0.032]{Mpc^{-1}}$. The vertical lines represent the time of recombination and that of matter-$\Lambda$ equality. The lower panels show the fractional difference between the numerical and analytical curves.
  }
	\label{fig:metric}
\end{figure}

\paragraph{Metric sources} \MHM computed the linear response of the photon perturbations to the second-order metric perturbations, which they obtained using an analytical approach in a radiation-free Universe.
The evolution of the second-order vector and tensor modes of the metric is approximately known for a standard $\Lambda$CDM Universe without radiation \cite{mollerach:2004a, boubekeur:2009a, matarrese:1998a}.
In \fref{fig:metric} we show a comparison between the numerical result from \SONG, which includes the effect of radiation, compared with the analytical formula in \MHM, for a standard $\Lambda$CDM cosmology and a medium-scale Fourier mode.
For the vector modes $\omega$ we use the $\Lambda\neq0$ approximation (Eq.~(10) of \MHM), while for the tensor modes $\gamma$ we use the simpler formula for a $\Lambda=0$ Universe (Eq.~(15) of \MHM), similarly to what is done in the B-mode analysis of \MHM.
As expected, the match is not accurate ($>10\%$ difference) around the time of recombination, when the fraction in energy density of radiation is still significant.
The agreement improves during the CDM-dominated era to reach the $0.1\%$ level, and it slightly worsens when the dark energy component becomes important. Today ($a=1$), the numerical and analytical curves match to the few-percent level.
We remark that when considering larger-scale Fourier modes we find an improved agreement at early times; as suggested by \MHM, this is due to the fact that the small-scale modes enter the horizon when the fraction of radiation is still significant, \ie when the analytical approximation fails.
Despite the approximate agreement for the second-order metric, our result seems to be in tension with that of \MHM, as the intrinsic B-mode power spectrum computed by \SONG, which includes also the effect computed in \MHM, is one order of magnitude smaller.
A similar tension was found by Schiffer \cite{schiffer:2007a}, who computed the linear response to the second-order metric perturbations to be a factor of seven smaller than the result of \MHM; in the same reference, it is claimed that the difference is mainly due to the evaluation of the integral over the visibility function going from Eq.~(49) to Eq.~(50) in \MHM.
Indeed, the authors of \MHM confirmed\footnote{S.~Matarrese, private communication.} that their computation of $\,C_\L^{BB}\,$ contained two numerical inaccuracies pertaining, respectively, to the width of the visibility function and to the ratio $\,\eta_0/\eta_D\,$ between the conformal time today and at decoupling, thus leading to a result approximately an order of magnitude too large.

\subsection{Convergence tests}

\label{sec:convergence}
Following \BF we test the numerical stability and convergence of our results by computing the power spectrum of the intrinsic B-modes at $\ell=500$ and $\ell=1500$ with varying numerical parameters; each parameter is varied while keeping all the others fixed.
Doing so allows us to explore the dependence of the final result on the numerical assumptions made in \SONG, and thus determine the optimal parameter values needed to obtain a target precision.
As a result of these convergence tests, we find that the results of \sref{sec:results} are stable at the $1\%$-level for all the relevant parameters.

\begin{figure}[t]
  \includegraphics[width=1\linewidth]{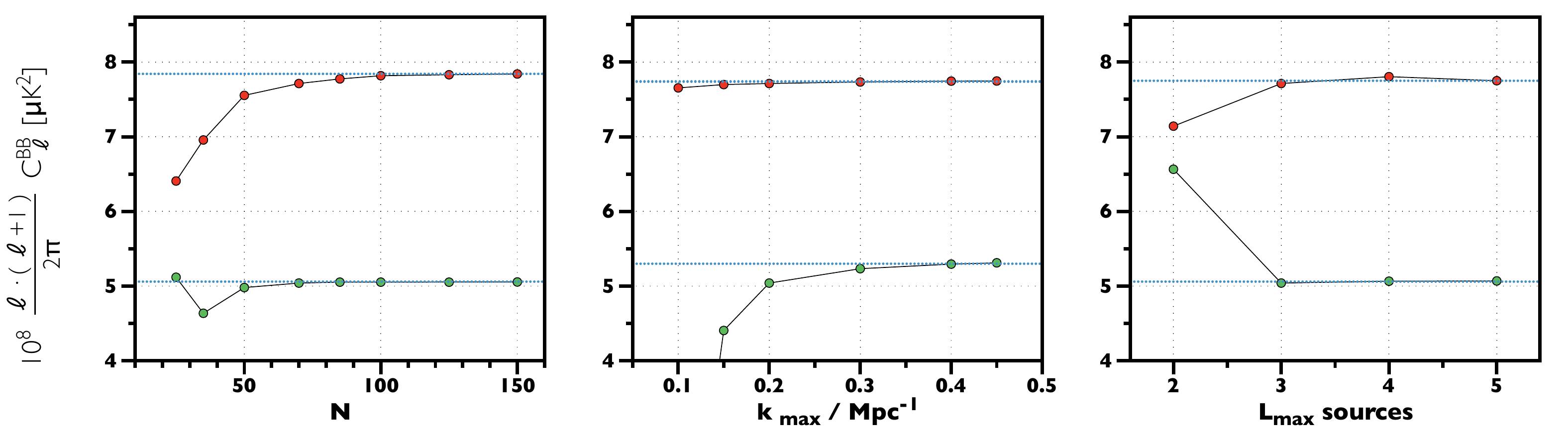}
  \caption{Numerical convergence of the vector contribution to $C_\ell^{BB}$, for $\ell=500$ (upper red points) and $\ell=1500$ (lower green points), for three numerical parameters: $\,k_{\text{max}}\,$, the maximum Fourier magnitude considered; $\,N\,$, the number of points in the Fourier and time grids; $\,L_{\text{max}}\,$, the highest multipole source considered in the line-of-sight integration. The horizontal dotted lines represent the value of the parameter at which the numerical result is stable at the $1\%$ level.}
\label{fig:vector_convergence}
\end{figure}


In \fref{fig:vector_convergence} we show the convergence of the three main numerical parameters relevant for B-polarisation.
First we consider the number of points $N$ in our grid in comoving wavevector $k$ and conformal time $\eta$. Increasing $N$ is numerically very expensive as the second-order differential system needs to be solved for approximately $N^3$ times. For the computation of our results we use $N=120$.

Next we show the convergence with respect to $k_\text{max}$, the maximum wavenumber considered. The power spectrum receives contributions from arbitrarily large wavevectors, but in practice these are strongly suppressed by the $k^{-3}$ dependence of the primordial power spectrum. Similar to the first-order case, we find that $k_\text{max}$ can be truncated to a few times $\lmax\cdot\eta_0$, where $\lmax$ is the maximum angular resolution needed ($\lmax=2000$ in our case) and $\eta_0$ is the conformal age of the Universe ($\eta_0\approx\unit[14]{Gpc}$ for a standard $\Lambda$CDM Universe). Our results presented in this paper use a value of $k_{\text{max}}=0.35\,\text{Mpc}^{-1}$.

Finally we investigate up to which multipole moment the second-order sources need to be considered. This number,  $L_\text{max}$, is the upper limit of the implicit summation over $\ell$ in the line-of-sight integral, \eref{eq:los_integral}. 
Because of the anisotropy suppression due to the tight-coupling regime at recombination, we expect the sources with $\ell>2$ to give progressively smaller contributions. 
We indeed find that we can safely choose $L_\text{max} = 3$, with the octupole still adding a significant contribution to the B-mode polarisation \cite{beneke:2011a}.

\section{Conclusions}
\label{sec:conclusions}

We have used the Boltzmann code \SONG \cite{pettinari:2013a} to compute the intrinsic B-mode polarisation induced in the CMB by the evolution of primordial density perturbations at second order. This intrinsic signal is completely determined by the standard cosmological parameters, is unavoidable and is not affected by the details of cosmic inflation. Our analysis extends the previous result from Beneke, Fidler and Klingm\"uller (2011)
\cite{beneke:2011a}, which focused on second-order scattering sources, 
by including the second-order metric sources, the quadratic Liouville terms at recombination and the non-linear transport 
terms related to the redshift effects in the Boltzmann hierarchy. The intrinsic B-polarisation is a guaranteed contribution to the B-mode power spectrum that exists even in the absence of primordial gravitational waves and might therefore contaminate their measurement.
It is complementary to other contributions such as from weak lensing \cite{zaldarriaga:1998c} and the time-delay effect \cite{hu:2001a}, which have already been calculated in the literature using non-perturbative approaches.
We have considered the second-order sources of B-polarisation consistently by taking into account their correlations, in contrast with the previous literature \cite{mollerach:2004a, beneke:2011a}, which focussed on specific limits and particular physical effects.
Furthermore, we have tested the robustness of our computations by performing extensive convergence tests on the numerical parameters of \SONG (\fref{fig:vector_convergence}), and by matching analytical limits for the vector and tensor modes of the metric at second order (\fref{fig:metric}).\\

At second order, many sources contribute to the B-polarisation. In this work, we have considered all those arising from the Boltzmann equation, excluding the well-known lensing and time-delay effects along the line-of-sight, which are better treated using non-perturbative approaches.
We have found that the major contribution to the intrinsic power spectrum comes from the sources at recombination (named $C_\ell^{\tilde{B}\tilde{B}}$ in \sref{sec:results}). These include the non-linear Compton collisions of the CMB photons off the electrons (\eref{eq:collision_term}) and the indirect effect of the non-scalar perturbations, which include metric perturbations (\eref{eq:metric_terms}), but also the quadratic Liouville terms (\eref{eq:QL}), which are computed in this paper for the first time.  The first two contributions have previously been separately estimated in the literature, while in our computation we consider them simultaneously and thus take into account their correlation.
When including only the collisional sources in \SONG, we agree with the results of Ref.~\cite{beneke:2011a}. The B-modes spectrum from these sources alone is twice as big as our full result (\fref{fig:compare_b_modes}), showing that there is a partial cancellation with the remaining effects considered.
On the other hand, we are not able to confirm the result of Mollerach, Harari \& Matarrese \cite{mollerach:2004a} (2004), as the intrinsic power spectrum computed by \SONG in an approximation similar to Ref.~\cite{mollerach:2004a} is one order of magnitude smaller. As discussed in \sref{sec:comparison_with_literature}, this discrepancy is merely numerical and is due to an inaccurate parametrisation of the visibility function and of the conformal-time ratio used in Ref.~\cite{mollerach:2004a}.\\

For the first time, we have treated the redshift contributions along the line-of-sight, extending the method proposed in Ref. \cite{huang:2013a} to the case of polarisation (\eref{eq:delta_tilde_transformation_2nd} and (\ref{eq:absorption})). This effect is the modification of the distribution function due to changes in the photon's momentum as it propagates through the inhomogeneous Universe, and needs to be integrated across the whole Hubble length.
We also express our results in terms of the bolometric temperature, $\Theta\,$, which we expect to be more closely related to observations than the brightness, $\Delta\,$ (\sref{sec:temperature_definition}). The B-modes in the bolometric temperature are connected to those in the brightness in a non-trivial way, which mixes E and B-polarisation.
We have included this novel contribution in our analysis, and found that the resulting B-modes partly cancel with those from the redshift contribution (\eref{eq:temperature_perturbation_redshift_plus_temperature}).
Considered separately, the redshift and bolometric-temperature corrections are comparable to  the recombination effects; together, they are subdominant.\\

After including all the effects described above, we have found that, at its peak ($\ell\approx700$), the power spectrum of the intrinsic B-polarisation is comparable to a primordial signal with a tensor-to-scalar ratio of $r\approx\sci{5}{-5}\,$ (\fref{fig:intrinsic_b_modes}).
The primordial signal, however, peaks on much larger angular scales ($\ell\approx100$), where the intrinsic spectrum is smaller and amounts to an equivalent $r$ of roughly $10^{-7}$.
This minimal overlap suggests that the measurements from future CMB experiments such as LiteBird \cite{hazumi:2012a,matsumura:2013a}, PIXIE \cite{kogut:2011a} and PRISM \cite{prism-collaboration:2013b}, with the latter claiming a sensitivity of $\Delta r=\sci{3}{-4}$, will not be biased by the intrinsic B-polarisation.
Therefore, the limiting factors for such experiment are likely to be the efficiency of the foreground-removal and de-lensing algorithms.
\paragraph{Note Added}
When we were completing this work, the BICEP2 collaboration \cite{bicep2-collaboration:2014a, bicep2-collaboration:2014b} has released the final results from the homonymous polarimeter, along with preliminary data from its successor, the Keck array \cite{sheehy:2010a}. Both data sets show an excess of power in the B-polarisation that is well fitted by a $\Lambda$CDM model with a tensor-to-scalar ratio of $\,r=0.20^{+0.07}_{-0.05}\,$, with $r=0$ disfavoured at $5.9\sigma\,$.
Planck's polarised maps and Keck's complete data are planned to be released by the end of 2014 and should provide a verification of this potentially far-reaching result.
If confirmed, such a high value of $r$ will make the study of primordial gravitational waves in the CMB less dependent on the contaminations discussed in this paper.

\acknowledgments
We wish to thank Christian Byrnes for an insightful discussion on the expected value of~$r$.
C.~Fidler, R.~Crittenden, K.~Koyama and D.~Wands are supported by the UK Science and Technology Facilities Council grants number ST/K00090/1 and ST/L005573/1, G.~Pettinari acknowledges support from grant number ST/I000976/1. The work of M.~Beneke is supported in part by the Gottfried Wilhelm 
Leibniz programme of the Deutsche Forschungsgemeinschaft (DFG).

\appendix
\section{Coupling coefficients}
\label{app:coupling}

In order to write the Boltzmann equation for the B-modes in a compact way (see Eqs.~\ref{eq:boltzmann_pure_bmodes} to \ref{eq:boltzmann_quad_collision_bmodes}), we have used the $D$ coefficients. These arise from the decomposition into spherical harmonics of the general Boltzmann equation \eref{eq:boltz}, and were introduced in Ref.~\cite{beneke:2010a}. They are defined as
\begin{align}
  &D^{\pm,\L}_{m_1\,m} \;\equiv\;
  \left(\begin{array}{cc|c}1&\L\pm1&\L \\ 0 & 2 & 2\end{array}\right)\,
  \left(\begin{array}{cc|c}1&\L\pm1&\L \\ m-m_1 & m_1 & m\end{array}\right) \nmsk
  &D^{0,\L}_{m_1\,m} \;\equiv\;
  -\left(\begin{array}{cc|c}1&\L&\L \\ 0 & 2 & 2\end{array}\right)\,
  \left(\begin{array}{cc|c}1&\L&\L \\ m-m_1 & m_1 & m\end{array}\right) \;,
  \label{eq:D_coupling_coefficients}
\end{align}
and they explicitly read
\begin{align}
  &D^{+,\L}_{m\pm1\,m} \;=\; -\frac{\sqrt{(\L-1)\:(\L+3)\:(\L+1\pm m)\:(\L+2\pm m)}}{\sqrt{2}\,(\L+1)\,(2\L+3)}\;,
  \;D^{-,\L}_{m\,m} \;=\;\frac{\sqrt{\L^2-4}}{\L}\;\frac{\sqrt{\L^2-m^2}}{2\L-1} \;,
  \nmsk
  &D^{+,\L}_{m\,m} \;=\;\frac{\sqrt{(\L-1)\:(\L+3)}}{\L+1}\;\frac{\sqrt{(\L+1)^2-m^2}}{2\L+3} \;,
  \;D^{-,\L}_{m\pm1\,m} \;=\; \frac{\sqrt{(\L^2-4)\:(\L-1\mp m)\:(\L\mp m)}}{\sqrt{2}\,\L\,(2\L-1)} \;,
  \nmsk
  &D^{0,\L}_{m\pm 1\,m} \;=\; \mp \frac{\sqrt{2(\L+1\pm m)\:(\L \mp m)}}{\L(\L+1)} \;,
  \;D^{0,\L}_{m\,m} \;=\; -\frac{2m}{\L (\L+1)} \;.
  \label{eq:coupling_coefficients}
\end{align}
The $D^0$ coefficients encode the mixing between the E and B modes, while the $D^{\pm}$ ones determine how the large $\L$-multipoles are generated from the lower ones via free-streaming.

\clearpage
\bibliographystyle{JHEP}
\bibliography{my_bibliography}
  
\end{document}